\documentclass[twocolumn, prl, showpacs]{revtex4}

\usepackage{epsfig}

\usepackage{amsmath}
\usepackage{amsfonts}
\usepackage{amssymb}

\begin{document}

\title{Spectral Coarse Graining and Synchronization in Oscillator Networks}

\author{David Gfeller$^{1,2}$} 
\author{Paolo De Los Rios$^2$}
\affiliation{$^1$Donnelly Centre for Cellular and Biomolecular Research, University of Toronto, 160 College Street, Toronto, Ontario, Canada M5S 3E1. \\ $^2$LBS, SB/ITP, Ecole Polytechnique F\'ed\'erale de Lausanne (EPFL), CH-1015, Lausanne, Switzerland.}

\date{\today} 

\begin{abstract}

Coarse graining techniques offer a promising alternative to large-scale simulations of complex dynamical systems, as long as the coarse-grained system is truly representative of the initial one. Here, we investigate how the dynamical properties of oscillator networks are affected when some nodes are merged together to form a coarse-grained network. Moreover, we show that there exists a way of grouping nodes preserving as much as possible some crucial aspects of the network dynamics. This coarse graining approach provides a useful method to simplify complex oscillator networks, and more generally any network whose dynamics involves a Laplacian matrix.

\end{abstract}

\pacs{05.45.Xt, 89.75.Fb}

\maketitle

Dynamical systems of coupled oscillators have been often used to describe several natural phenomena in various fields of science ranging from cardiology to ecology~\cite{Strogatz2003}. In particular, the emergence of synchronization has received much attention. Historically, the first results have been obtained considering some particular system architecture~\cite{Kuramoto1975, Pecora1998}. More recently, synchronization has been studied for arbitrary coupling given by the topology of complex networks~\cite{Barahona2002, Nishikawa2003, Chavez2005, Motter2005-2, Arenas2006}.

Unfortunately, for large dynamical systems of interacting units such as cardiac cells synchronizing in the heart or Malaysian fireflies flashing in unison~\cite{Strogatz2003}, considering the full network often results in a very large number of coupled differential equations. In such cases, the use of coarse-grained models is key to reducing the system complexity, and a crucial issue is to know how we should formulate the coarse-grained model and its dynamics~\cite{Gonzalez2007}.

A common approach is to consider some units of the system as almost indistinguishable and to merge them into one single node, giving rise to the concept of meta-populations~\cite{Colizza2007}. Most often, prior information about the nature of the nodes has been used to decide which ones should be merged. However, some attempts can be found to define automated procedures depending only on the information given by the network itself. In particular, it has been shown recently that one can merge nodes so that the properties of random walks on the network are almost left unchanged~\cite{Gaveau2005, Lafon2006, Gfeller2007-2}. Mathematically, this was expressed as the possibility of preserving some eigenvalues of the stochastic matrix. Merging nodes is also related to the problem of finding clusters in networks (see for instance~\cite{Girvan2002, Reichardt2004, Danon2005} for popular algorithms, or~\cite{Oh2005, Arenas2006, Lodato2007} for methods based on network synchronization). However, the first aim of clustering techniques is rather to classify the nodes into communities and not to build a coarse-grained model of a network.

In this Letter, we investigate how the dynamics of coupled oscillator networks, and especially synchronization patterns, is affected by merging some nodes together. In particular, we show that the method of~\cite{Gfeller2007-2}, referred to as Spectral Coarse Graining (SCG), can be extended to provide a natural framework for coarse graining oscillator networks preserving some of their dynamical properties.

We consider a system of $N$ identical oscillators with a coupling given by the topology of an undirected network. Most often the coupling involves the Laplacian $L$, which is defined as $L_{ij}=-w_{ji}$ if $i\neq j$ and $L_{ii}=\sum_{j\neq i}L_{ij}$, where $w_{ji}$ stands for the weight of the edge from node $j$ to node $i$. We briefly recall that all rows of $L$ sum up to 0 ($\sum_jL_{ij}=0$), which implies that there is an eigenvalue $\lambda^1=0$ with a corresponding constant eigenvector $p^1$. Moreover, for undirected and connected networks, the eigenvalues of $L$ are all real and satisfy $0=\lambda^1< \lambda^2 \leq  \hdots \leq \lambda^N$. These eigenvalues play a critical role in the dynamics of oscillator networks, as we will see below.

Our first example of oscillator network dynamics falls in the general framework of~\cite{Pecora1998}:
\begin{equation}
\dot{{\bf x}}_i=F({\bf x}_i)+\sigma\sum_{j=1}^NL_{ij}H({\bf x}_j),
\label{sync.eq}
\end{equation}
with ${\bf x}_i \in \mathbb{R}^d$. $F({\bf x})$ accounts for the internal dynamics of each node, $H({\bf x})$ is a coupling function and $\sigma$ is the coupling strength. 
In this case, the eigenvalues of $L$ appear naturally in the analysis of the stability of the synchronized state (${\bf x}_i(t)={\bf s}(t)$, $\forall i$, $\forall t$), as it was shown in the seminal work of Ref.~\cite{Pecora1998}. There, the authors have proved that the linear stability of this state is described by the variational equation $\dot{\xi}_i=DF({\bf s})\xi_i+\sigma\sum_{j=1}^NL_{ij}DH({\bf s})\xi_i$, which can be diagonalized into: 
\[\dot{\zeta}^\alpha=DF({\bf s})\zeta^\alpha+\sigma\lambda^\alpha DH({\bf s})\zeta^\alpha, \hspace{5mm} \alpha=1,\hdots, N.\]
The synchronized state ($\alpha=1$) is linearly stable if all Lyapunov exponents are negative for $\alpha=2,\hdots, N$. Interestingly, in several cases, such as the R\"ossler oscillators~\cite{Rossler1976}, there exists a single range of values $l_1\leq\sigma\lambda^\alpha \leq l_2$, $\alpha=2, \hdots ,N$ such that the synchronized state is linearly stable. As a consequence, the network can be synchronized if and only if $\lambda^N/\lambda^2<l^2/l^1=\beta$~\cite{Pecora1998, Barahona2002}.

Other dynamical processes have also been investigated to study synchronization in networks, such as the Kuramoto model~\cite{Kuramoto1975}:
\begin{equation}
\dot{x}_i=\omega_i+\sigma \sum_jA_{ij}\sin(x_j-x_i).
\label{kuramoto.eq}
\end{equation}
Here $A_{ij}$ stands for the adjacency matrix of the network. Close to the synchronized state ($x_i=x_j$) and considering identical oscillators ($\omega_i=\omega$, $\forall i$) as in~\cite{Arenas2006}, the linearized dynamics is given by $\dot{\bf{x}}=\Omega-\sigma L\bf{x}$, with $\Omega=(\omega_1, \hdots, \omega_N)^T$. In this case the synchronized state is always stable and the eigenvalues of $L$ describe the behavior of the normal modes of the linearized dynamics~\cite{Arenas2006}. 

Overall, the two dynamical systems described above show that the eigenvalues of $L$ play a crucial role in the dynamics of oscillator networks. It is therefore extremely important to know how these eigenvalues are affected when some nodes are merged. Besides, one should try to preserve their value as much as possible. The coarse graining strategy introduced in this Letter shows that this goal can be reached in many cases.

\begin{figure}[]
\begin{center}
\includegraphics[width=80mm]{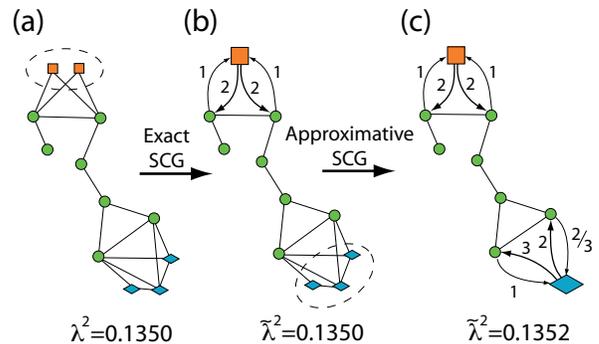}
  \caption{(Color online) (a) Small toy network. (b) Coarse-grained network in which the two square nodes have been merged. (c) Coarse-grained network in which the three diamond nodes have further been merged. Numbers on the edges indicate the weight of the edges that have been updated in the coarse graining. The smallest non-trivial eigenvalue of $L$, $\lambda^2$, is displayed for each network.}
  \label{exact.fig}
\end{center}
\end{figure}

To define properly a coarse graining strategy, two questions need to be answered.
First, how should we merge nodes and update the edges in oscillator networks, so that the resulting network is truly representative of the initial one? Second, which nodes should be merged? The first question, though less important if one is only interested in identifying groups of nodes in a network, is crucial to build a coarse-grained network in which each node has similar dynamical properties as the initial nodes it is made of. For simplicity, we first consider the example of the small network of Figure~\ref{exact.fig}~(a). Let us assume we have decided to merge the two square nodes. Then the edges in the reduced network should be drawn as in  Figure~\ref{exact.fig}~(b). Indeed, since each of both circle nodes connected to the square nodes in (a) receives two edges from them, an edge with weight equal to two needs to be drawn in the reduced network. On the other hand, we want the square node in the reduced network of (b) to exhibit a behavior corresponding to the one of the two square nodes in the initial network. Since each of these two nodes receives only one edge from each circle node, an edge with weight equal to one is drawn in the network of Figure~\ref{exact.fig}~(b).

More generally, the weight of the edges should first be added for each group. Then, to preserve the structure of equation~(\ref{sync.eq}) or~(\ref{kuramoto.eq}) the weight of incoming edges should be divided by the size of the group they are pointing to, as in Figure~\ref{exact.fig}~(b)~and~(c). 
Interestingly, if the resulting network is made up of $\tilde{N}$ nodes corresponding to $\tilde{N}$ groups in the initial network, the Laplacian $\tilde{L}$ of the reduced network can be expressed as a matrix product
\begin{equation}
\tilde{L}=KLR.
\label{KR1.eq}	
\end{equation}
$K\in \mathbb{R}^{\tilde{N}\times N}$ and $R\in \mathbb{R}^{N\times \tilde{N}}$ are two rectangular matrices defined as follows. If nodes are labeled with $i=1,\hdots N$ and groups are labeled with $C=1,\hdots, \tilde{N}$, then
\begin{equation}
  K_{Ci}=\delta_{C,C_i}\frac{1}{|C|} \hspace{2mm} \text{and} \hspace{2mm} R_{Ci}=\delta_{C,C_i},
  \label{KR2.eq}
\end{equation}
where $|C|$ is the cardinality of group $C$, $C_i$ is the label of the group of node $i$, and $\delta$ is the usual Kronecker symbol. Therefore, the $C^{\text{th}}$ row of $R$ (column of $K$) has non-zero elements only for the entries corresponding to the nodes in group $C$. Equations~(\ref{KR1.eq}) and~(\ref{KR2.eq}) define the central matrix transformation for the Spectral Coarse Graining (SCG) of $L$. We note that the rows of $\tilde{L}$ sum up to zero for any kind of grouping, which means that $\tilde{L}$ always has the properties of a Laplacian. Having found how edges are to be updated when nodes are merged, we still have to know which nodes should be merged. 

This question can be answered by observing that equations~(\ref{KR1.eq}) is very similar the one used in Ref.~\cite{Gfeller2007-2} for the SCG of stochastic matrices ~\footnote{Actually they become equivalent if one takes the transpose of (\ref{KR2.eq}), which has no effect on the eigenvalues}.
In particular, we have the property that $RK$ is block diagonal (up to an appropriate reshuffling of the rows and columns) with each block corresponding to a group of nodes, and that all entries within a block are equal to the inverse of the group size. 
This means that if groups are formed such that the components of an eigenvector $p^\alpha$ of $L$ are equal within each group, then $RKp^\alpha=p^\alpha$, which implies that the reduced vector $Kp^\alpha$ is an eigenvector of $\tilde{L}$ with eigenvalue $\lambda^\alpha$ ($\tilde{L}Kp^\alpha=KLRKp^\alpha=\lambda^\alpha Kp^\alpha$). Therefore, merging nodes with the same components in $p^\alpha$ preserves the eigenvalue $\lambda^\alpha$ of $L$, i.e. $\lambda^\alpha$ is also an eigenvalue of $\tilde{L}$. This property defines the {\em exact} SCG of $L$.
The equality between eigenvector components is for instance encountered if some nodes have exactly the same neighbors, as the two square nodes in Figure~\ref{exact.fig}~(a)~\footnote{In this case, all eigenvalues are preserved in the network of Figure~\ref{exact.fig}~(b) except $\lambda^\beta=L_{11}+L_{12}$, for which one can show that $p_1^\beta+p_2^\beta=0$ and $p_i^\beta=0$, $\forall i>2$.}.

Furthermore, the perturbation approach outlined in~\cite{Gfeller2007-2} for stochastic matrices applies as well to the SCG of $L$. If two nodes $i$ and $j$ having almost the same eigenvector components in $p^\alpha$ are merged, the new Laplacian $\tilde{L}$ has an eigenvalue $\tilde{\lambda}^{\tilde{\alpha}}$ almost equal to $\lambda^\alpha$. In this case, the SCG is said to be {\em approximate}. Mathematically, the fact nodes $i$ and $j$ have almost the same eigenvector components can be expressed for instance as $d^\alpha_{i,j}\equiv |p^\alpha_i - p^\alpha_j|/(p^\alpha_{\text{max}}-p^\alpha_{\text{min}}) \ll 1$ for a given $\alpha$, where $p^\alpha_{\text{max}}$ ($p^\alpha_{\text{min}}$) are the largest (smallest) components of $p^\alpha$. In practice, the condition $d_{ij}^\alpha\ll 1$ can be implemented by defining $I$ equally distributed intervals between $p^\alpha_{\text{min}}$ and $p^\alpha_{\text{max}}$ and grouping the nodes whose eigenvector components in $p^\alpha$ fall in the same interval. 

In Figure~\ref{exact.fig}, we provide an example of the two kinds of SCG considering the eigenvector $p^2$. On the one hand, the two square nodes in Figure~(\ref{exact.fig}) have exactly the same eigenvector components in $p^2$ and the eigenvalue $\lambda^2$ is exactly preserved in the network of Figure~\ref{exact.fig}~(b). On the other hand, the components in $p^2$ of the three nodes shown with blue diamonds in Figure~\ref{exact.fig}~(b) are very close to each other (all $d^2_{i,j}$ are smaller than 0.02) and the eigenvalue $\lambda^2$ is almost preserved in the network of Figure~\ref{exact.fig}~(c).
This shows that SCG is indeed not restricted to the special case in which eigenvector components are equal, but applies also when components are close to each other. Finally, we note that SCG can be readily extended to preserve more than one eigenvalue by merging nodes that have almost the same components within each corresponding eigenvector.

\begin{figure}[t] 
\begin{center}
\includegraphics[width=87mm]{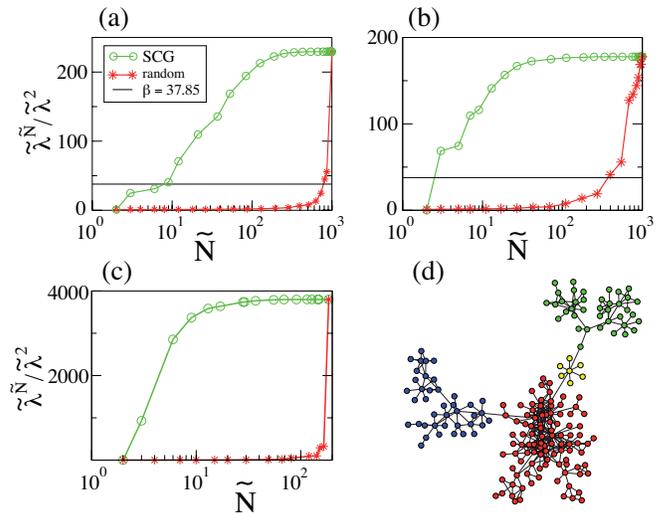}
  \caption{(Color online) (a)-(c) Evolution of the ratio $\tilde{\lambda}^{\tilde{N}}/\tilde{\lambda}^2$ as a function of the size of the coarse-grained network. Red stars correspond to a random merging of the nodes; green circles correspond to the SCG using both $p^2$ and $p^N$. The black line shows the critical value $\beta=37.85$. (a) SW network~\cite{Watts1998}, $N=1000$, $p=0.05$. (b) BA network~\cite{Barabasi1999}, $N=1005$, $m=2$. (c) Network displayed in (d). (d) Sub-component of a synonymy network ($N$=185). Colors were set only to help for comparison with Figure~\ref{kuramoto.fig}~(a).}
  \label{teigev.fig}
\end{center}
\end{figure}

To illustrate the effect of SCG on the synchronization patterns in large oscillator networks, we first consider the dynamics of R\"ossler oscillators~\cite{Rossler1976} with $x$-coupling [$F(x,y,z)=\left(-y-z, x+ay, b+z(x-c)\right)$, with $a=b=0.2$ and $c=7$, $H(x,y,z)=x$]. In this case, the stability of the synchronized state is given by the condition $\lambda^N/\lambda^2<\beta\approx 37.85$~\cite{Pecora1998}. 
The behavior of $\lambda^N/\lambda^2$ has already been studied in several kinds of networks~\cite{Barahona2002, Nishikawa2003, Chavez2005}. Here, instead, we investigate how $\lambda^N/\lambda^2$ changes when some nodes are merged. 
In Figure~\ref{teigev.fig}~(a)~and~(b), we first show our results for two generic networks: a Barab\'asi-Albert (BA)~\cite{Barabasi1999} and a small-world (SW) network~\cite{Watts1998}.
Stars correspond to a random merging of the nodes into $\tilde{N}$ groups, while circles show the results of SCG performed along both $p^2$ and $p^N$, i.e, groups are made up of nodes with $d_{ij}^2\ll1$ and $d_{ij}^N\ll1$. In the latter case, the different values of $\tilde{N}$ correspond to different choices of the number of intervals $I$ defined between the smallest and the largest component in the eigenvectors $p^2$ and $p^N$ (the smaller $I$, the smaller $\tilde{N}$). Both eigenvalues $\lambda^2$ and $\lambda^N$ are well preserved in the network obtained with SCG even when the size is strongly reduced. 
A similar analysis was carried out in Figure~\ref{teigev.fig}~(c) for a real network of $N=185$ nodes diplayed in Figure~\ref{teigev.fig}~(d). The network was originally built from a synonymy relationship between words and the graph displayed in Figure~\ref{teigev.fig}~(d) corresponds to one of the disconnected sub-components~\cite{Gfeller2005}. This network exhibits clearly a heterogeneous internal structure, which makes it particularly interesting to illustrate a coarse graining procedure.

Figure~\ref{teigev.fig} clearly shows  that, if nodes are merged randomly, the ratio $\lambda^2$ and $\lambda^N$ changes dramatically. On the other hand, the nature of the synchronized state is preserved much longer if groups are formed using the SCG along both $p^2$ and $p^N$.

\begin{figure}[t] 
\begin{center}
\includegraphics[width=87mm]{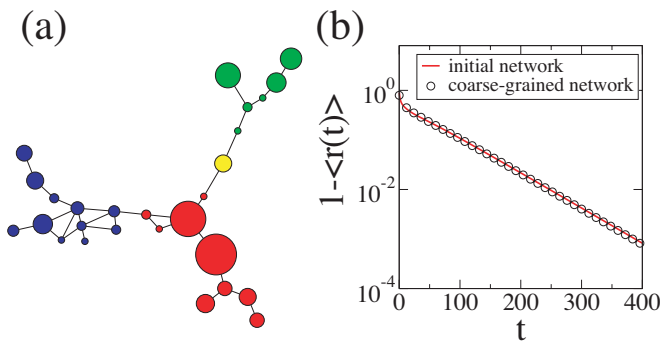}
  \caption{(Color online) (a) Coarse-grained version ($\tilde{N}$=27) of the network in Figure~\ref{teigev.fig}~(d). The node size is proportional to the size of the corresponding groups. The node color in the coarse-grained network corresponds to the color of the nodes in the initial network (in Figure~\ref{teigev.fig}~(d) colors were chosen such that each group is made up of nodes of only one color).  (b) Temporal evolution of $1-\langle r(t)\rangle$ according to equation~\ref{kuramoto.eq} in the initial network of Figure~\ref{teigev.fig} (continuous line) and in the coarse-grained network displayed in (a) (circles). $\sigma=1$.}
  \label{kuramoto.fig}
\end{center}
\end{figure}

Let us now consider the Kuramoto model of equation~(\ref{kuramoto.eq}). We have seen that the lowest eigenvalues of $L$ describe the slow modes of the system close to the synchronized state. Thus one should try to preserve these eigenvalues in a coarse-grained network. In Figure~\ref{kuramoto.fig}~(a), we have built a coarse-grained version the network of Figure~\ref{teigev.fig}~(d). This coarse-grained network was obtained by SCG considering the two lowest non-zero eigenvalues of $L$, $\lambda^2=0.0081$ and $\lambda^3=0.016$. Along each of the corresponding eigenvectors $p^2$ and $p^3$, we have defined $I=40$ intervals equally distributed between the smallest and largest eigenvector component.  We have grouped nodes that fall in the same interval for each eigenvector.

The two eigenvalues of $L$ are well preserved in the new Laplacian $\tilde{L}$: $\tilde{\lambda}^2=0.0083$ and $\tilde{\lambda}^3=0.017$. To compare the dynamics defined by the non-linearized equation~(\ref{kuramoto.eq}) in the two networks, we display in Figure~\ref{kuramoto.fig}~(b) the temporal evolution of the order parameter~\cite{Kuramoto1984}: 
$1-\langle r(t)\rangle=1-\langle\frac{1}{N}\left[\left(\sum_i\cos(x_i(t))\right)^2+\left(\sum_i\sin(x_i(t))\right)^2\right]^{1/2}\rangle,$
where the average is taken over a set of 200 randomly chosen initial conditions. The synchronized state (given by $1-r(t)=0$) is reached in a very similar way in both networks, which allows to conclude that the coarse-grained network is representative of the initial one from the point of view of the synchronization. We also observe that in this case, the coarse-grained network of Figure~\ref{kuramoto.fig}~(a) can be interpreted as a backbone structure of the initial network.

Summarizing, we have investigated how synchronization in oscillator networks behaves when some nodes are merged together. For this purpose, we have introduced a Spectral Coarse Graining technique for the Laplacian matrix, which allowed us to know both {\em how} nodes should be merged, and {\em which} nodes should be merged in oscillator networks. This technique requires only the computation of a few selected eigenvalues and eigenvectors, and was shown to preserve some crucial aspects of the network dynamics. For instance in the case of R\"ossler oscillators, the key parameter is the ratio $\lambda^N/\lambda^2$ and our results show that SCG allows to reduce significantly the network size while keeping $\lambda^N/\lambda^2$ close to its initial value, whereas  random merging of the nodes had dramatics effects on $\lambda^N/\lambda^2$. Considering the dynamics of the Kuramoto model, we have shown that the behavior of the system close to the synchronized state can be preserved by preserving the first eigenvalues of $L$.
More generally, we believe that SCG provides an appropriate mathematical framework to simplify a variety of dynamical systems whose dynamics involves a coupling given by the Laplacian of a network.

Finally, with a view to enhancing synchronizability~\cite{Barahona2002, Nishikawa2003, Chavez2005} instead of preserving the dynamics, SCG may be useful to decide which nodes should be merged such that the ratio $\tilde{\lambda}^{\tilde{N}}/\tilde{\lambda}^2$ displays the steepest decrease with $\tilde{N}$.

We thank D. Morton de Lachapelle and M. Angeles Serrano for a critical reading of the manuscript. This work was financially supported by COSIN (NET Open IST 2001-33555), DELIS (FET Open 001907), the SER-Bern (02.0234) and the Swiss National Science Foundation (Grant PBELA--120936) .


\begin{thebibliography}{23}
\expandafter\ifx\csname natexlab\endcsname\relax\def\natexlab#1{#1}\fi
\expandafter\ifx\csname bibnamefont\endcsname\relax
  \def\bibnamefont#1{#1}\fi
\expandafter\ifx\csname bibfnamefont\endcsname\relax
  \def\bibfnamefont#1{#1}\fi
\expandafter\ifx\csname citenamefont\endcsname\relax
  \def\citenamefont#1{#1}\fi
\expandafter\ifx\csname url\endcsname\relax
  \def\url#1{\texttt{#1}}\fi
\expandafter\ifx\csname urlprefix\endcsname\relax\def\urlprefix{URL }\fi
\providecommand{\bibinfo}[2]{#2}
\providecommand{\eprint}[2][]{\url{#2}}

\bibitem[{\citenamefont{Strogatz}(2003)}]{Strogatz2003}
\bibinfo{author}{\bibfnamefont{S.~H.} \bibnamefont{Strogatz}},
  \emph{\bibinfo{title}{Sync: The Emerging Science of Spontaneous Order}}
  (\bibinfo{publisher}{Hyperion Press}, \bibinfo{year}{2003}).

\bibitem[{\citenamefont{Kuramoto}(1975)}]{Kuramoto1975}
\bibinfo{author}{\bibfnamefont{Y.}~\bibnamefont{Kuramoto}}, in
  \emph{\bibinfo{booktitle}{Lecture Notes in Physics}}
  (\bibinfo{publisher}{Springer}, \bibinfo{address}{New York},
  \bibinfo{year}{1975}), vol.~\bibinfo{volume}{39}, p. \bibinfo{pages}{420}.

\bibitem[{\citenamefont{Pecora and Carroll}(1998)}]{Pecora1998}
\bibinfo{author}{\bibfnamefont{L.~M.} \bibnamefont{Pecora}} \bibnamefont{and}
  \bibinfo{author}{\bibfnamefont{T.~L.} \bibnamefont{Carroll}},
  \bibinfo{journal}{Phys. Rev. Lett.} \textbf{\bibinfo{volume}{80}},
  \bibinfo{pages}{2109} (\bibinfo{year}{1998}).

\bibitem[{\citenamefont{Barahona and Pecora}(2002)}]{Barahona2002}
\bibinfo{author}{\bibfnamefont{M.}~\bibnamefont{Barahona}} \bibnamefont{and}
  \bibinfo{author}{\bibfnamefont{L.~M.} \bibnamefont{Pecora}},
  \bibinfo{journal}{Phys. Rev. Lett.} \textbf{\bibinfo{volume}{89}},
  \bibinfo{pages}{054101} (\bibinfo{year}{2002}).

\bibitem[{\citenamefont{Nishikawa et~al.}(2003)\citenamefont{Nishikawa, Motter,
  Lai, and Hoppensteadt}}]{Nishikawa2003}
\bibinfo{author}{\bibfnamefont{T.}~\bibnamefont{Nishikawa}},
  \bibinfo{author}{\bibfnamefont{A.~E.} \bibnamefont{Motter}},
  \bibinfo{author}{\bibfnamefont{Y.}~\bibnamefont{Lai}}, \bibnamefont{and}
  \bibinfo{author}{\bibfnamefont{F.~C.} \bibnamefont{Hoppensteadt}},
  \bibinfo{journal}{Phys. Rev. Lett.} \textbf{\bibinfo{volume}{91}},
  \bibinfo{pages}{014101} (\bibinfo{year}{2003}).

\bibitem[{\citenamefont{Chavez et~al.}(2005)\citenamefont{Chavez, Hwang, Amann,
  Hentschel, and Boccaletti}}]{Chavez2005}
\bibinfo{author}{\bibfnamefont{M.}~\bibnamefont{Chavez}},
  \bibinfo{author}{\bibfnamefont{D.-U.} \bibnamefont{Hwang}},
  \bibinfo{author}{\bibfnamefont{A.}~\bibnamefont{Amann}},
  \bibinfo{author}{\bibfnamefont{H.~G.~E.} \bibnamefont{Hentschel}},
  \bibnamefont{and}
  \bibinfo{author}{\bibfnamefont{S.}~\bibnamefont{Boccaletti}},
  \bibinfo{journal}{Phys. Rev. Lett.} \textbf{\bibinfo{volume}{94}},
  \bibinfo{pages}{218701} (\bibinfo{year}{2005}).

\bibitem[{\citenamefont{Motter et~al.}(2005)\citenamefont{Motter, Zhou, and
  Kurths}}]{Motter2005-2}
\bibinfo{author}{\bibfnamefont{A.~E.} \bibnamefont{Motter}},
  \bibinfo{author}{\bibfnamefont{C.}~\bibnamefont{Zhou}}, \bibnamefont{and}
  \bibinfo{author}{\bibfnamefont{J.}~\bibnamefont{Kurths}},
  \bibinfo{journal}{Europhysics Letters} \textbf{\bibinfo{volume}{69}},
  \bibinfo{pages}{334} (\bibinfo{year}{2005}).

\bibitem[{\citenamefont{Arenas et~al.}(2006)\citenamefont{Arenas,
  D\'iaz-Guilera, and P\'erez-Vicente}}]{Arenas2006}
\bibinfo{author}{\bibfnamefont{A.}~\bibnamefont{Arenas}},
  \bibinfo{author}{\bibfnamefont{A.}~\bibnamefont{D\'iaz-Guilera}},
  \bibnamefont{and} \bibinfo{author}{\bibfnamefont{C.~J.}
  \bibnamefont{P\'erez-Vicente}}, \bibinfo{journal}{Phys. Rev. Lett.}
  \textbf{\bibinfo{volume}{96}}, \bibinfo{pages}{114102}
  (\bibinfo{year}{2006}).

\bibitem[{\citenamefont{Gonzalez and Barab\'asi}(2007)}]{Gonzalez2007}
\bibinfo{author}{\bibfnamefont{M.~C.} \bibnamefont{Gonzalez}} \bibnamefont{and}
  \bibinfo{author}{\bibfnamefont{A.-L.} \bibnamefont{Barab\'asi}},
  \bibinfo{journal}{Nature} \textbf{\bibinfo{volume}{3}}, \bibinfo{pages}{224}
  (\bibinfo{year}{2007}).

\bibitem[{\citenamefont{Colizza et~al.}(2007)\citenamefont{Colizza,
  Pastor-Satorras, and Vespignani}}]{Colizza2007}
\bibinfo{author}{\bibfnamefont{V.}~\bibnamefont{Colizza}},
  \bibinfo{author}{\bibfnamefont{R.}~\bibnamefont{Pastor-Satorras}},
  \bibnamefont{and}
  \bibinfo{author}{\bibfnamefont{A.}~\bibnamefont{Vespignani}},
  \bibinfo{journal}{Nature Physics} \textbf{\bibinfo{volume}{3}},
  \bibinfo{pages}{276} (\bibinfo{year}{2007}).

\bibitem[{\citenamefont{Gaveau and Schulman}(2005)}]{Gaveau2005}
\bibinfo{author}{\bibfnamefont{B.}~\bibnamefont{Gaveau}} \bibnamefont{and}
  \bibinfo{author}{\bibfnamefont{L.~S.} \bibnamefont{Schulman}},
  \bibinfo{journal}{Bulletin des Sciences Math\'ematiques}
  \textbf{\bibinfo{volume}{129}}, \bibinfo{pages}{631} (\bibinfo{year}{2005}).

\bibitem[{\citenamefont{Lafon and Lee}(2006)}]{Lafon2006}
\bibinfo{author}{\bibfnamefont{S.}~\bibnamefont{Lafon}} \bibnamefont{and}
  \bibinfo{author}{\bibfnamefont{A.~B.} \bibnamefont{Lee}},
  \bibinfo{journal}{IEEE Transactions on Pattern Analysis and Machine
  Intelligence} \textbf{\bibinfo{volume}{28}}, \bibinfo{pages}{1393}
  (\bibinfo{year}{2006}).

\bibitem[{\citenamefont{Gfeller and {De Los Rios}}(2007)}]{Gfeller2007-2}
\bibinfo{author}{\bibfnamefont{D.}~\bibnamefont{Gfeller}} \bibnamefont{and}
  \bibinfo{author}{\bibfnamefont{P.}~\bibnamefont{{De Los Rios}}},
  \bibinfo{journal}{Phys. Rev. Lett.} \textbf{\bibinfo{volume}{99}},
  \bibinfo{pages}{038701} (\bibinfo{year}{2007}).

\bibitem[{\citenamefont{Girvan and Newman}(2002)}]{Girvan2002}
\bibinfo{author}{\bibfnamefont{M.}~\bibnamefont{Girvan}} \bibnamefont{and}
  \bibinfo{author}{\bibfnamefont{M.~E.~J.} \bibnamefont{Newman}},
  \bibinfo{journal}{PNAS} \textbf{\bibinfo{volume}{99}}, \bibinfo{pages}{7821}
  (\bibinfo{year}{2002}).

\bibitem[{\citenamefont{Reichardt and Bornholdt}(2004)}]{Reichardt2004}
\bibinfo{author}{\bibfnamefont{J.}~\bibnamefont{Reichardt}} \bibnamefont{and}
  \bibinfo{author}{\bibfnamefont{S.}~\bibnamefont{Bornholdt}},
  \bibinfo{journal}{Phys. Rev. Lett.} \textbf{\bibinfo{volume}{93}},
  \bibinfo{pages}{218701} (\bibinfo{year}{2004}).

\bibitem[{\citenamefont{Danon et~al.}(2005)\citenamefont{Danon, Diaz-Guilera,
  Duch, and Arenas}}]{Danon2005}
\bibinfo{author}{\bibfnamefont{L.}~\bibnamefont{Danon}},
  \bibinfo{author}{\bibfnamefont{A.}~\bibnamefont{Diaz-Guilera}},
  \bibinfo{author}{\bibfnamefont{J.}~\bibnamefont{Duch}}, \bibnamefont{and}
  \bibinfo{author}{\bibfnamefont{A.}~\bibnamefont{Arenas}},
  \bibinfo{journal}{J. Stat. Mech.} p. \bibinfo{pages}{09008}
  (\bibinfo{year}{2005}).

\bibitem[{\citenamefont{Oh et~al.}(2005)\citenamefont{Oh, Rho, Hong, and
  Kahng}}]{Oh2005}
\bibinfo{author}{\bibfnamefont{E.}~\bibnamefont{Oh}},
  \bibinfo{author}{\bibfnamefont{K.}~\bibnamefont{Rho}},
  \bibinfo{author}{\bibfnamefont{H.}~\bibnamefont{Hong}}, \bibnamefont{and}
  \bibinfo{author}{\bibfnamefont{B.}~\bibnamefont{Kahng}},
  \bibinfo{journal}{Phys. Rev. E} \textbf{\bibinfo{volume}{72}},
  \bibinfo{pages}{047101} (\bibinfo{year}{2005}).

\bibitem[{\citenamefont{Lodato et~al.}(2007)\citenamefont{Lodato, Boccaletti,
  and Latora}}]{Lodato2007}
\bibinfo{author}{\bibfnamefont{I.}~\bibnamefont{Lodato}},
  \bibinfo{author}{\bibfnamefont{V.}~\bibnamefont{Boccaletti}},
  \bibnamefont{and} \bibinfo{author}{\bibfnamefont{V.}~\bibnamefont{Latora}},
  \bibinfo{journal}{Europhysics Letters} \textbf{\bibinfo{volume}{78}},
  \bibinfo{pages}{28001} (\bibinfo{year}{2007}).

\bibitem[{\citenamefont{R\"ossler}(1976)}]{Rossler1976}
\bibinfo{author}{\bibfnamefont{O.~E.} \bibnamefont{R\"ossler}},
  \bibinfo{journal}{Phys. Lett.} \textbf{\bibinfo{volume}{57}},
  \bibinfo{pages}{397} (\bibinfo{year}{1976}).

\bibitem[{\citenamefont{Barab\'asi and Albert}(1999)}]{Barabasi1999}
\bibinfo{author}{\bibfnamefont{A.-L.} \bibnamefont{Barab\'asi}}
  \bibnamefont{and} \bibinfo{author}{\bibfnamefont{R.}~\bibnamefont{Albert}},
  \bibinfo{journal}{Science} \textbf{\bibinfo{volume}{286}},
  \bibinfo{pages}{509} (\bibinfo{year}{1999}).

\bibitem[{\citenamefont{Watts and Strogatz}(1998)}]{Watts1998}
\bibinfo{author}{\bibfnamefont{D.~J.} \bibnamefont{Watts}} \bibnamefont{and}
  \bibinfo{author}{\bibfnamefont{S.~H.} \bibnamefont{Strogatz}},
  \bibinfo{journal}{Nature} \textbf{\bibinfo{volume}{393}},
  \bibinfo{pages}{440} (\bibinfo{year}{1998}).

\bibitem[{\citenamefont{Gfeller et~al.}(2005)\citenamefont{Gfeller, Chappelier,
  and {De Los Rios}}}]{Gfeller2005}
\bibinfo{author}{\bibfnamefont{D.}~\bibnamefont{Gfeller}},
  \bibinfo{author}{\bibfnamefont{J.-C.} \bibnamefont{Chappelier}},
  \bibnamefont{and} \bibinfo{author}{\bibfnamefont{P.}~\bibnamefont{{De Los
  Rios}}}, \bibinfo{journal}{Phys. Rev. E} \textbf{\bibinfo{volume}{72}},
  \bibinfo{pages}{056135} (\bibinfo{year}{2005}).

\bibitem[{\citenamefont{Kuramoto}(1984)}]{Kuramoto1984}
\bibinfo{author}{\bibfnamefont{Y.}~\bibnamefont{Kuramoto}},
  \emph{\bibinfo{title}{Chemical Oscillations, Waves and Turbulence}}
  (\bibinfo{publisher}{Springer}, \bibinfo{address}{New York},
  \bibinfo{year}{1984}).

\end{thebibliography}
\end{document}